\documentclass[
aps,
twocolumn,
prapplied,
superscriptaddress,
letterpaper,
nofootinbib,
longbibliography]{revtex4-2}
\usepackage{amsmath}
\usepackage{amssymb}
\usepackage{physics}
\usepackage{mathtools}
\usepackage{upgreek}
\usepackage{booktabs}
\usepackage{lipsum}
\usepackage[justification=Justified,format=plain]{caption}
\usepackage{bm}
\usepackage{tabularx}
\usepackage{booktabs}
\usepackage[colorlinks=true,
            linkcolor=blue,
            citecolor=blue,
            urlcolor=blue]{hyperref}
\AtBeginDocument{
	\heavyrulewidth=.08em
	\lightrulewidth=.05em
	\cmidrulewidth=.03em
	\belowrulesep=.65ex
	\belowbottomsep=0pt 
	\aboverulesep=.4ex
	\abovetopsep=0pt
	\cmidrulesep=\doublerulesep
	\cmidrulekern=.5em
	\defaultaddspace=.5em
}

\begin{document}

\title{Fundamental limitations of absolute ranging via\\deep frequency modulation interferometry}

\author{Miguel Dovale-\'Alvarez}
\affiliation{James C.\ Wyant College of Optical Sciences, University of Arizona, 1630 E.\ University Blvd., Tucson, Arizona 85721, USA}

\begin{abstract}
Deep frequency modulation interferometry (DFMI) resolves phase ambiguity in absolute distance measurements by jointly estimating two length-encoding parameters: the coarse and unambiguous effective modulation depth ($m$), and the fine but ambiguous interferometric phase ($\phi$). We establish a comprehensive framework quantifying the fundamental precision limits and practical accuracy constraints of this technique. A Fisher-information analysis defines the intrinsic estimator precision for $m$ and $\phi$, while the contribution of carrier frequency drift introduces an additional, time-dependent source of random error. Numerical simulations reveal a structured error landscape with previously unrecognized ``valleys of robustness,'' where systematic biases from common hardware imperfections are suppressed by orders of magnitude. An analytical model based on signal orthogonality explains their origin and predicts their locations. The results yield a consolidated error budget accounting for both random and systematic errors, providing a quantitative design paradigm for absolute length metrology via DFMI.
\end{abstract}

\maketitle

\section{Introduction} \label{section:introduction}

Precision length metrology is foundational to a wide range of scientific and technological domains, from gravitational-wave detection to advanced manufacturing~\cite{Estler2002}. In space-based science, missions like the Laser Interferometer Space Antenna (LISA)~\cite{LISA, LISARedBook} require absolute ranging to synthesize interferometers over million-kilometer baselines~\cite{Tinto2005, Tinto1999, Yamamoto2024, Reinhardt2024}. Similarly, Earth-observing missions such as GRACE Follow-On~\cite{Sheard2012, GRACEFO} and NGGM~\cite{Nicklaus2020, Nicklaus2022} depend on absolute distance measurements to map Earth's gravity field. This paradigm is central to future missions involving precision formation flying for astronomy or geodesy~\cite{Zhang2022, Lawson2006, Fridlund2008, Cash2000, Turyshev2007, Turyshev2009}. This need is mirrored in terrestrial applications, where absolute metrology is indispensable for advanced manufacturing and quality control in industries ranging from semiconductor manufacturing to aerospace engineering~\cite{deGroot2019, Huang2025, Novak2003, Everton2016, Chen2024, FraunhoferILT2025}.

Conventional interferometry offers exceptional sensitivity but suffers from an inherent ambiguity~\cite{Jacobs1981}. Its output, the phase $\phi$, is ``wrapped'' into a single $2\pi$ cycle. The absolute optical path length $\Delta l$ is related to this measurement by an unknown integer fringe order, $N$:
\begin{equation}
\Delta l = \left( N + \frac{\phi}{2\pi} \right) \lambda_0 = \frac{\Phi}{2\pi} \lambda_0,
\label{eq:abs_length_reconstruction}
\end{equation}
where $\lambda_0$ is the laser wavelength, $\phi \in [0, 2\pi)$ is the observable interferometric phase, and $\Phi = 2\pi \Delta l / \lambda_0$ is the unwrapped interferometric phase.

Deep frequency modulation interferometry (DFMI) resolves this ambiguity by impressing a strong sinusoidal frequency modulation onto the laser~\cite{Gerberding2015, Kissinger2015, Heinzel2010}. The resulting signal encodes $\Delta l$ into two parameters: the precise but ambiguous interferometric phase,
\begin{equation}
\phi = \frac{2\pi f_0 \Delta l}{c} \pmod{2 \pi},
\end{equation}
and the coarse but unambiguous modulation depth,
\begin{equation}
m = \frac{2\pi \Delta f \Delta l}{c}.
\end{equation}
Here, $f_0$ is the laser carrier frequency and $\Delta f$ is the modulation amplitude. An accurate measurement of $m$ provides a coarse estimate of $\Delta l$ sufficient to determine $N$, thereby enabling an absolute measurement with the precision of $\phi$.

Extracting these parameters requires sophisticated readout algorithms, typically either frequency-domain batch processors like the nonlinear least squares (NLS) fit~\cite{Schwarze2014, Isleif2016, Isleif2019} or time-domain sequential estimators such as the extended Kalman filter (EKF)~\cite{Vorndamme2017MSc}. While the statistical precision limit for estimating $\phi$ has been established~\cite{Eckhardt2022}, a complete framework for quantifying the fundamental and practical limits of absolute length metrology via DFMI has been lacking.

This paper establishes such a comprehensive framework. We first derive the Cram\'er--Rao Lower Bound (CRLB) for the estimation of the effective modulation depth $m$---the key parameter governing ambiguity resolution---and demonstrate that both the NLS and EKF estimators asymptotically achieve this limit. We then analyze the impact of laser carrier frequency drift as the other principal contributor to random error. Extending the analysis to systematic effects, we develop a digital-twin model of a DFMI experiment and perform extensive simulations to quantify the biases introduced by modulation nonlinearity, residual amplitude modulation, and ghost beams. This investigation reveals the existence of ``valleys of robustness''---specific operating points where these biases are intrinsically minimized. Their origin and locations are derived from first principles using a signal orthogonality framework. Finally, we synthesize these results into a consolidated error budget that incorporates random noise, systematic effects, and scale-factor calibration errors, thereby clarifying the fundamental trade-offs that govern high-accuracy DFMI system design.

The reminder of this paper is structured as follows. Section~\ref{sec:DFMI_Signal} details the DFMI signal model and the principle of absolute length reconstruction. Section~\ref{sec:DFMI_Readout} introduces the readout algorithms used to estimate $m$ and $\phi$. Section~\ref{sec:Statistical_Precision} establishes the fundamental statistical limits of the technique, governed by fundamental estimator precision and laser carrier frequency drift. Section~\ref{sec:Systematic_Limitations} analyzes the primary systematic error sources arising from hardware imperfections and scale factor calibration errors. Section~\ref{sec:results} synthesizes these results into a consolidated error budget for fringe-resolved absolute length determination. Finally, Section~\ref{sec:conclusion} concludes the paper. Supplementary derivations are provided in the appendices.

\section{DFMI Signal Model and Ranging Principle}
\label{sec:DFMI_Signal}

\begin{figure}
\centering
\includegraphics[trim={0mm 5mm 0mm 0mm},clip,width=\columnwidth]{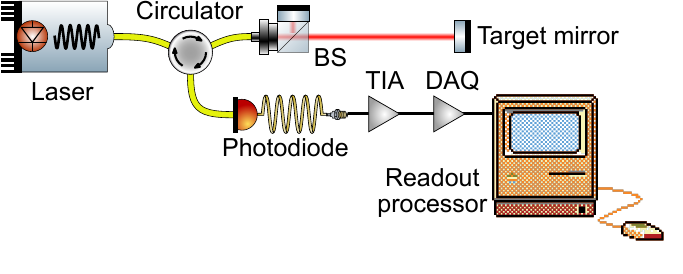}
\caption{A typical DFMI setup based on a Michelson interferometer. A laser with sinusoidal frequency modulation is split into a reference arm and a measurement arm. The recombined light is detected by a photodiode, and the resulting voltage signal is digitized and processed. The setup is adapted from~\cite{Rohr2023}.}
\label{fig:setup}
\end{figure}

\begin{figure}
\centering
\includegraphics{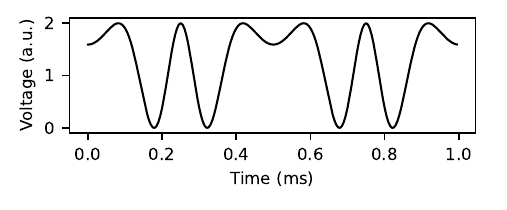}
\caption{Example DFMI signal for an interferometer with $\Delta l = 5\,\rm cm$, injected with a laser modulated by $6.87\,\rm GHz$ at $1\,\rm kHz$. The resulting configuration yields a modulation depth $m = 7.2\,\rm rad$.}
\label{fig:DFMI-signal}
\end{figure}

In a DFMI system (Figure~\ref{fig:setup}), a laser's frequency is sinusoidally modulated around a carrier $f_0$. The resulting instantaneous frequency deviation is $f_{\text{mod}}(t) = \Delta f \cos(2 \pi f_m t + \psi)$. When this light is injected into an interferometer with an optical path difference $\Delta l$, the time delay $\tau = \Delta l/c$ between the arms produces a total detected phase difference. Under the common condition $2 \pi f_m \tau \ll 1$, justified in Section~\ref{appendix:taylor_error}, the photodetector voltage signal (Figure~\ref{fig:DFMI-signal}) can be modeled as
\begin{equation}
v(t) = A + C \cdot \cos\left(\Phi + m \cos \left( 2 \pi f_m t + \psi \right) \right).
\label{eq:volt_signal}
\end{equation}
Here, $A$ is a DC offset, $C$ is the signal's AC amplitude, and the key parameters $\Phi$ and $m$ are the interferometric phase and modulation depth defined in Section~\ref{section:introduction}.

This periodic signal can be analyzed in the frequency domain. Using the Jacobi-Anger expansion, the complex amplitude of the signal's $n$-th harmonic ($n \geq 0$) can be modeled as
\begin{equation}
a_n(\mathbf{x}) = 2C \cdot J_n(m) \cos\left(\Phi + n\frac{\pi}{2}\right) e^{-in\psi},
\label{eq:alpha_n}
\end{equation}
where $\mathbf{x} = (\Phi, \psi, m, C)^{\intercal}$ is the vector of core interferometric parameters, and $J_n(m)$ is the $n$-th order Bessel function of the first kind. The real (in-phase) and imaginary (quadrature) parts of this complex amplitude, $a_n = I_n + iQ_n$, are explicitly given by
\begin{align}
I_n(\mathbf{x}) &= 2C \cdot J_n(m) \cos\left(\Phi + n\frac{\pi}{2}\right) \cos(n\psi), \label{eq:Iana} \\
Q_n(\mathbf{x}) &= -2C \cdot J_n(m) \cos\left(\Phi + n\frac{\pi}{2}\right) \sin(n\psi). \label{eq:Qana}
\end{align}
The total DC voltage measured is $V_{\text{DC}} = A + a_0(\mathbf{x})/2 = A + C J_0(m)\cos(\Phi)$. Since the instrumental offset $A$ is independent of the other parameters, frequency-domain estimators typically discard the DC component and fit the model only to the AC harmonics ($n \geq 1$) to determine $\mathbf{x}$. This model shows how the parameters are uniquely encoded in the signal's AC harmonic structure.

Note that, because the signal in Equation~\eqref{eq:volt_signal} is $2\pi$-periodic in~$\Phi$, any estimator operating on this model recovers only the wrapped phase, $\hat{\phi} = \hat{\Phi} \bmod 2\pi$.

Absolute length reconstruction in DFMI is a three-step process. First, the unambiguous modulation depth estimate, $\hat{m}$, is used to form a coarse but absolute estimate of the interferometric phase:
\begin{equation}
\hat \Phi_{\text{coarse}} = \hat{m} \frac{f_{0, \rm est}}{\Delta f_{\rm est}}.
\label{eq:coarse_phase}
\end{equation}
Here, $f_{0, \rm est}$ and $\Delta f_{\rm est}$ are the calibrated estimates of the laser frequency and tuning amplitude, respectively. 

Second, this coarse phase is used to determine the integer fringe order $N$ of the precise, wrapped phase estimate $\hat \phi$ by finding the closest integer:
\begin{equation}
\hat{N} = \text{round}\left( \frac{\hat \Phi_{\text{coarse}} - \hat \phi}{2\pi} \right).
\label{eq:N_estimate}
\end{equation}
Lastly, the estimates $\hat N$ and $\hat \phi$ are combined in Equation~\eqref{eq:abs_length_reconstruction} to form an absolute measurement of $\Delta l$ with the precision of $\hat \phi$.

The rounding operation in Equation~\eqref{eq:N_estimate} succeeds if the total error between the coarse estimate and the true phase is less than half a fringe. To guarantee high-confidence resolution, a robust design criterion must account for the worst-case combination of systematic biases ($b$) and random errors ($\sigma$):
\begin{equation}
|b_{\rm coarse}| + |b_{\phi}| + K \sqrt{\sigma_{\rm coarse}^2 + \sigma_{\phi}^2} < \pi.
\label{eq:ambiguity_condition_comprehensive_intro}
\end{equation}
Here, $b_{\rm coarse}$ and $b_{\phi}$ are the biases in the $\hat \Phi_{\text{coarse}}$ and  $\hat \phi$ estimators, respectively, $\sigma_{\rm coarse}^2$ and $\sigma_{\phi}^2$ are their variances, and $K$ is a confidence factor (e.g., $K=3$ for a 99.7\% probability bound on the random error for a Gaussian process). This inequality ensures that even in the presence of the maximum expected systematic bias, the probability of a random fluctuation causing a fringe-order error is negligible. 

It is important to clarify the statistical model underlying this condition. The coarse phase estimate, $\hat \Phi_{\text{coarse}}$, is a continuous-valued quantity derived from the measurement of the modulation depth, $\hat{m}$. Because $\hat{m}$ and $\hat{\phi}$ are the result of an estimation process (e.g., NLS or EKF) on a noisy signal, their error distribution is asymptotically Gaussian, a standard result for maximum-likelihood-type estimators in the high signal-to-noise ratio regime. Consequently, the error on the coarse phase is also treated as a Gaussian random variable.

\section{DFMI Readout Algorithms}
\label{sec:DFMI_Readout}

\begin{figure*}
\centering
\includegraphics{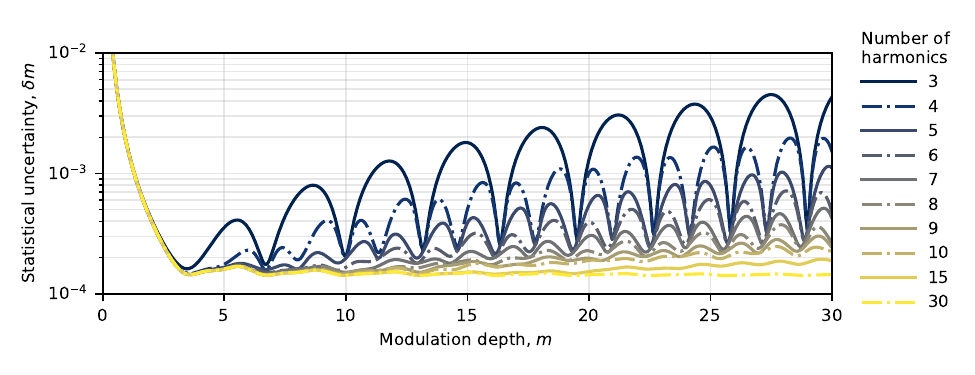}
\caption{Statistical uncertainty of the modulation depth estimate ($\sigma_m$) as a function of the true modulation depth $m$ and the number of harmonics ($N_{h}$) used in the fit. The curves show the Cram\'er-Rao Lower Bound, calculated from the inverse of the full $4\times 4$ FIM, assuming a fixed raw SNR of $80\,\rm dB$. For low $N_{h}$, the precision exhibits sharp degradation in ``dead zones''. As more harmonics are included, these zones disappear and the precision becomes a smoother function of $m$, converging to an asymptotic limit.}
\label{fig:dm_vs_ndata}
\end{figure*}

The system parameters are estimated from the digitized signal using specialized readout algorithms. Two complementary approaches are employed: a frequency-domain nonlinear least-squares (NLS) fit, implemented as a batch estimator, and a time-domain extended Kalman filter (EKF), implemented as a sequential estimator. Both algorithms are available in the open-source python package \texttt{DeepFMKit}~\cite{Dovale2025}.

Both readout schemes estimate the parameter vector
$\mathbf{x} = (\Phi, \psi, m, C)^{\intercal}$.
Because the signal model in Equation~\eqref{eq:volt_signal} is $2\pi$-periodic in~$\Phi$, the recovered phase component $\hat{\Phi}$ is inherently wrapped to~$[0, 2\pi)$ and thus corresponds to the observable phase $\hat{\phi}$ defined in Section~\ref{section:introduction}.

The NLS algorithm operates on the signal’s AC harmonics. It first demodulates the time-domain signal to extract the measured complex amplitudes of the first $N_h$ harmonics, denoted $\alpha_n$ for $n \in [1, N_h]$. It then determines the parameter set $\hat{\mathbf{x}}$ that minimizes the sum of squared residuals between these measured amplitudes and the analytical model of Equation~\eqref{eq:alpha_n}:
\begin{equation}
\label{eq:ssq}
S(\mathbf{x}) = \sum_{n=1}^{N_h} \left| a_{n}(\mathbf{x}) - \alpha_n \right|^2.
\end{equation}

In contrast, the EKF operates directly on the time-domain voltage samples, estimating the full five-parameter state vector
$\mathbf{x}_{\text{EKF}} = (\Phi, \psi, m, C, A)^{\intercal}$,
where $A$ is the DC offset. The process model typically assumes stationary parameters with a random-walk prior,
$\mathbf{x}_{\text{EKF},k+1} = \mathbf{x}_{\text{EKF},k} + \mathbf{w}_k$,
and the nonlinear measurement model
\begin{equation}
v_k = h(\mathbf{x}_{\text{EKF},k}, t_k) + \nu_k,
\end{equation}
relates the state to the measured sample $v_k$, with $h$ defined by Equation~\eqref{eq:volt_signal}.
As discussed in Section~\ref{sec:Statistical_Precision}, both estimators are asymptotically efficient and can approach the Cram\'er--Rao Lower Bound.

\section{Fundamental Statistical Limitations}
\label{sec:Statistical_Precision}

\subsection{The Cram\'er-Rao Lower Bound}
\label{sec:CRLB}

The statistical properties of the estimators introduced in Section~\ref{sec:DFMI_Readout} are governed by the Fisher Information~\cite{Kay1993} of the parameter vector $\mathbf{x} = (\Phi, \psi, m, C)^{\intercal}$. The performance of any unbiased estimator is ultimately constrained by the Cram\'er-Rao Lower Bound (CRLB), which provides a quantitative limit on precision based on signal properties and noise statistics. While the CRLB for the interferometric phase $\Phi$ has been established~\cite{Eckhardt2022}, the corresponding limit for the effective modulation depth $m$---the key to absolute length determination---has not. This section derives the CRLB for $m$, analyzes its dependence on system parameters, and establishes an asymptotic limit.

The CRLB is derived from the Fisher Information Matrix (FIM). For the frequency-domain DFMI model with additive white Gaussian noise in the time domain, the FIM is given by $\mathcal{I}(\mathbf{x}) = (\mathbf{J}^\mathrm{H}\mathbf{J})/\sigma_{h}^2$, where $\mathbf{J}$ is the Jacobian of the signal model~\cite{Kay1993}. The term $\sigma_{h}^2$ is the noise variance on the demodulated complex harmonic amplitudes. If these amplitudes are obtained via a Discrete Fourier Transform of a block of $N_v$ voltage samples, this variance is related to the time-domain voltage noise variance $\sigma_v^2$ by $\sigma_{h}^2 = 2\sigma_v^2/N_v$. The minimum achievable variance for any parameter is found from the inverse of the FIM, $\mathrm{Var}(\hat{m}) \ge \left[ \mathcal{I}^{-1}(\mathbf{x}) \right]_{m,m}$.

Numerical evaluation of this bound (Figure~\ref{fig:dm_vs_ndata}) reveals a complex dependence on the system's operating point and the number of harmonics ($N_{h}$) used in the estimation. For models incorporating very few harmonics, the precision degrades sharply in certain ``dead zones.'' These zones arise at values of $m$ where the small set of available harmonic amplitudes is insufficient to fully disentangle the four coupled parameters $(\Phi, \psi, m, C)$, leading to an ill-conditioned Fisher Information Matrix. However, for more realistic models where a larger number of harmonics is used ($N_{h} \ge 10$), information is distributed across many harmonic components. The loss of information from any single harmonic approaching a null becomes non-critical, the FIM remains well-conditioned, and the pronounced dead zones disappear. The precision then becomes a smoother function of $m$ that improves with the inclusion of more harmonics, eventually converging to an asymptotic limit that is independent of $m$.

\begin{figure}
\centering
\includegraphics[width=\columnwidth]{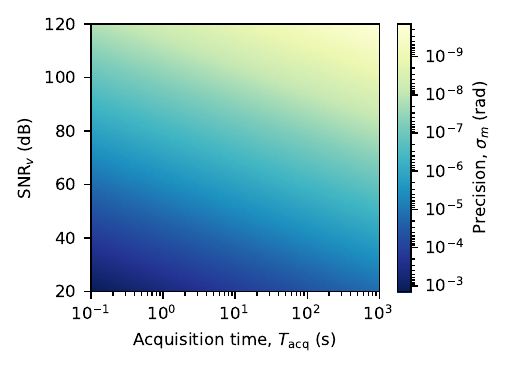}
\caption{The trade-off between the raw voltage signal-to-noise ratio SNR$_v$ and the acquisition time $T_{\rm acq}$ required to meet a target precision for the modulation depth estimate, $\hat m$. The plot illustrates the design trade-space for a system sampled at $f_s = 200\,\rm kHz$, where the total number of samples is $N_v = T_{\rm acq} \cdot f_s$.}
\label{fig:CRLB_SNR_Tacq}
\end{figure}

In the limit where a large number of harmonics is used, the CRLB converges to an asymptotic value that can be expressed more simply in terms of the time-domain signal properties. As detailed in Appendix~\ref{sec:CRLB_derivation}, the asymptotic CRLB for the variance of $\hat{m}$ is
\begin{equation}
\sigma_m^2 \approx \frac{4}{N_{v} \cdot \text{SNR}_{v}^2}.
\label{eq:CRLB_voltage}
\end{equation}
This result shows that the ultimate precision is governed by two fundamental quantities: the total number of samples acquired, $N_{v}$, and the raw signal-to-noise ratio of those samples, $\text{SNR}_{v} = C/\sigma_v$. Thus, for a given system, achieving a certain precision demands a minimum combination of signal quality and acquisition time, a trade-off illustrated in Figure~\ref{fig:CRLB_SNR_Tacq}.

Together with the asymptotic CRLB for the interferometric phase [Equation~\eqref{eq:CRLB_phi_voltage}], these results define the fundamental statistical limits of the technique. A further source of randomness in any practical implementation is the drift of the laser’s carrier frequency, which is addressed in the following section.

\subsection{Carrier frequency drift}
\label{sec:frequency-drift}

The statistical limit derived in Section~\ref{sec:CRLB} reveals that the precision of the modulation depth estimate $\hat{m}$ improves with longer acquisition times ($T_{\rm acq}$). However, a competing effect arises from the intrinsic drift of the laser carrier frequency, $f_0(t)$, which introduces a phase error that worsens over time. This creates a fundamental trade-off, as the total statistical uncertainty for ambiguity resolution does not indefinitely decrease with integration time.

We model the laser carrier frequency drift as a random walk, a process characteristic of free-running lasers, with a power spectral density (PSD) of $S_{f}(f) = A_{\rm f, RW}^2 f^{-2}$. This frequency noise couples into the interferometric signal via the time delay $\tau = \Delta l/c$, producing a differential phase wander, $\Phi_{\rm drift}(t)$. The variance of this phase wander grows linearly with observation time, meaning its standard deviation scales as:
\begin{equation}
\sigma_{\rm drift} \approx 2\pi A_{\rm f, RW} \tau \sqrt{T_{\rm acq}}.
\label{eq:phase_noise_random_walk}
\end{equation}
This growing uncertainty in the phase can, for long baselines or acquisition times, become a dominant error source that compromises ambiguity resolution on its own.

Crucially, the estimations of $\Phi$ and $m$ are affected very differently by this drift. The drift is a direct, cumulative error on the interferometric phase measurement $\hat \phi$, which tracks the physical reality of the laser's random walk. In contrast, the modulation depth $m$ is encoded in the relative amplitudes of the high-frequency signal harmonics at $f_m, 2f_m, \dots$. Due to a separation of timescales, the slow, low-frequency phase wander has negligible spectral power at these high harmonic frequencies. The demodulation process used by the NLS algorithm effectively acts as a high-pass filter, strongly rejecting the influence of the drift on the measured harmonic amplitudes. Consequently, the resulting error on $\hat{m}$ is extremely small and, importantly, does not accumulate over time.

This reveals a critical trade-off for system design: while longer acquisition times reduce the statistical uncertainty in $\hat{m}$ (improving $\hat \Phi_{\rm coarse}$), they increase the drift-induced uncertainty in the interferometric phase measurement (degrading $\hat \phi$). In long-baseline interferometry, this drift can become the dominant statistical limitation to robust ambiguity resolution.

\section{Systematic errors in DFMI}
\label{sec:Systematic_Limitations}

While statistical noise limits precision, systematic errors are the dominant barrier to achieving high accuracy in absolute length determination. Unlike the random error $\sigma_{\rm coarse}$, that can be reduced by longer integration, systematic errors introduce a persistent bias that often sets a hard limit on performance. This section analyzes the impact of modulation nonlinearity, residual amplitude modulation, and ghost beams, the primary sources of systematic bias in DFMI experiments.

Throughout this section, $\Phi$ and $m$ denote the true interferometric phase and effective modulation depth, while $\hat{\phi}$ and $\hat{m}$ refer to their respective estimates obtained from the NLS readout algorithm.
    
\subsection{Simulation framework}
\label{subsec:framework}

All analyses in this section are carried out through numerical simulations using a digital-twin model of a DFMI experiment. The model is implemented using the open-source python package \texttt{DeepFMKit}~\cite{Dovale2025}, a comprehensive framework for end-to-end simulation and analysis of DFMI systems. This approach enables controlled and quantitative exploration of complex error mechanisms that are often intractable to capture with purely analytical methods.

At the core of this framework lies a high-fidelity physics engine that extends beyond the idealized signal model of Equation~\eqref{eq:volt_signal}. It accurately reproduces the complete signal generation process, including arbitrary laser modulation waveforms, dynamic time-of-flight delays resolved via high-order interpolation, and the injection of realistic colored noise sources. The methodology for analyzing a given systematic effect follows a two-step procedure:

\paragraph{Signal generation with known imperfections.} The physics engine is configured to generate a realistic time-domain voltage signal that incorporates a specific, known hardware non-ideality. For example, to study modulation nonlinearity, the ideal sinusoidal modulation waveform is replaced with a distorted version containing a calibrated amount of second-harmonic content.

\paragraph{Analysis with an idealized model.} This generated, ``imperfect'' signal is then processed by the standard NLS readout algorithm described in Section~\ref{sec:DFMI_Readout}. Crucially, the NLS fitter is configured with the ideal signal model (Equation~\eqref{eq:alpha_n}), as this reflects the situation in a real experiment where the readout algorithm is unaware of the hardware's imperfections. The discrepancy between the parameters recovered by the NLS fit and the known ground-truth parameters of the simulation constitutes the systematic bias.

By performing this two-step process in large-scale Monte Carlo simulations that sweep across the parameter space and average over unknown phases, we can precisely map the magnitude of the systematic bias as a function of the system's operating point, as shown in the following subsections.

\subsection{Modulation nonlinearity}
\label{sec:NL}

\begin{figure*}
\centering
\includegraphics[width=\textwidth]{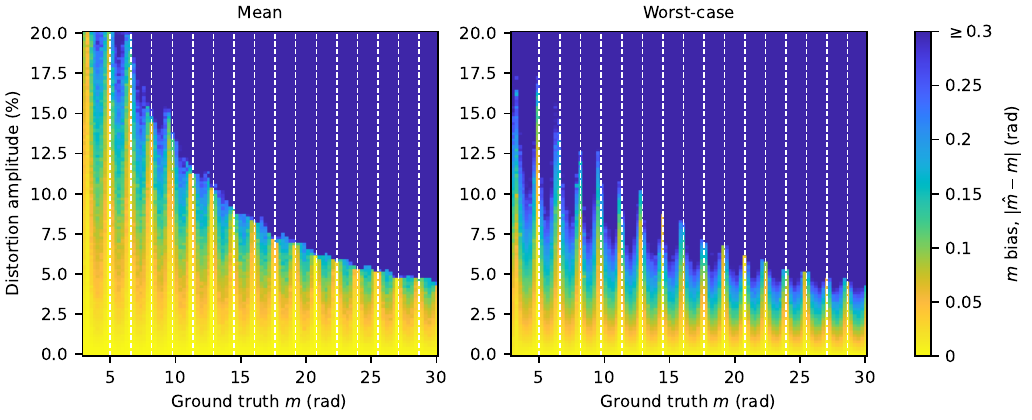}
\includegraphics[width=\textwidth]{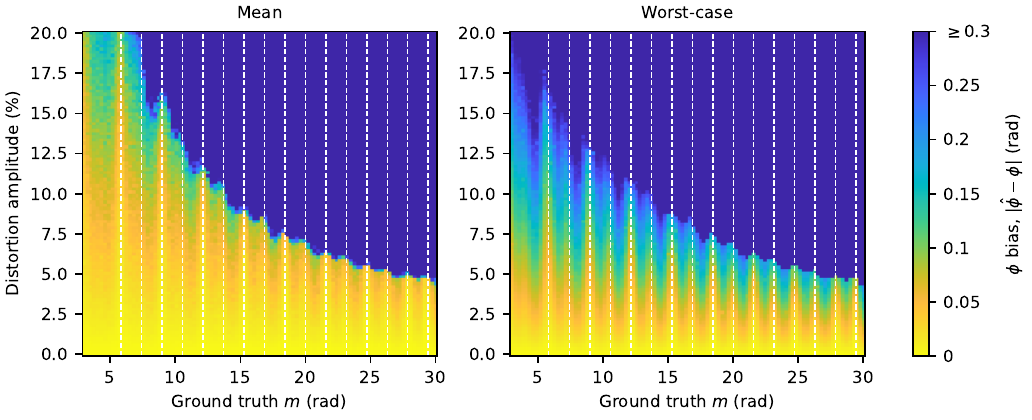}
\caption{Bias in the estimated modulation depth ($\hat{m}$) and interferometric phase ($\hat \phi$) due to second-harmonic distortion in the frequency modulation. The bias is shown as a function of the true modulation depth $m$ and the distortion amplitude $\epsilon$. The left and right panels show the mean and worst-case bias, respectively, from a Monte Carlo simulation over the unknown phases $\Phi$ and $\psi_2$. The plots reveal distinct vertical ``valleys of robustness'' where the bias is strongly suppressed. The $\hat m$ valleys align with the extrema of $J_2(2m)$, i.e., occurring at $m \approx 3.35, 4.98, 6.59, 8.17 \dots$, and the $\hat \phi$ valleys align with the zeros of $J_2(2m)$ starting at $m \approx 5.81$, i.e., occurring at $m \approx 5.81, 7.40, 8.98 \dots$. The dashed white lines indicate the theoretical predictions.}
\label{fig:distortion_m}
\end{figure*}

An ideal DFMI system assumes a purely sinusoidal laser frequency modulation. In practice, modulators often exhibit harmonic distortion, which violates this assumption and biases the estimated parameters. We model the most common form of this distortion by including a second-harmonic component in the instantaneous frequency modulation:
\begin{equation}
    f_{\rm mod}(t) = \Delta f \left[ \cos(2 \pi f_m t + \psi) + \epsilon \cos(4 \pi f_m t + \psi_2) \right],
\label{eq:f_mod_distorted}
\end{equation}
where $\epsilon \ll 1$ is the fractional amplitude of the distortion and $\psi_2$ is its phase. This distortion introduces a parasitic term into the interferometric phase, which for small perturbations results in a corresponding perturbation to the measured voltage signal:
\begin{align}
\delta v(t) \approx -C m \epsilon &\cos(4 \pi f_m t + \psi_2) \\ \nonumber 
\cdot &\sin(\Phi + m\cos(2 \pi f_m t + \psi)).
\label{eq:nl_perturbation_signal}	
\end{align}
This error signal reveals a mixing process: the $2 f_m$ distortion term beats against the ideal DFMI signal's harmonic structure. This corrupts the measured amplitudes of the signal's harmonics, thereby biasing the parameters estimated from them.

While the bias scales linearly with the distortion amplitude $\epsilon$, its dependence on the modulation depth $m$ is highly non-uniform. As shown by the simulations in Figure~\ref{fig:distortion_m}, there exist distinct vertical ``valleys of robustness''---specific operating points where the bias in $\hat{m}$ and $\hat \phi$ is suppressed by orders of magnitude.

The origin of these valleys can be explained by a signal orthogonality principle. A parameter estimate is robust to a signal perturbation if that perturbation is, on average, orthogonal to the signal's sensitivity with respect to that parameter. The sensitivity is given by the gradient of the ideal signal model, $g_x(t) = \partial v / \partial x$. The condition for minimal bias in a parameter $x$ is therefore $\langle \delta v(t), g_x(t) \rangle = 0$, where the angle brackets denote a time-average over a modulation period.

For the modulation depth $m$, the gradient is $g_m(t) = -C \cos(2 \pi f_m t + \psi) \sin(\Phi + m\cos(2 \pi f_m t + \psi))$. The orthogonality condition $\langle \delta v, g_m \rangle = 0$, when averaged over all unknown phases, requires that the derivative of the second-order Bessel function, $J'_2(2m)$, must be zero (see Appendix~\ref{appendix:orthogonality_derivation}). This condition predicts that robustness valleys for $\hat{m}$ occur at the extrema of $J_2(2m)$, a result in excellent agreement with the simulations.

Conversely, for the interferometric phase, the gradient is $g_\Phi(t) = -C \sin(\Phi + m\cos(2 \pi f_m t + \psi))$. The corresponding orthogonality condition requires that $J_2(2m)=0$. This prediction also aligns perfectly with the simulated results and highlights a critical design trade-off: the operating points that minimize bias for $\hat{m}$ are distinct from those that do so for $\hat \phi$.

\subsubsection*{Mitigation strategies}

Beyond selecting these robust operating points, two primary strategies exist for mitigating modulation nonlinearity.
\paragraph{Hardware compensation.} The most direct approach is to improve the modulator's linearity. This can involve using an arbitrary waveform generator (AWG) to apply a pre-distorted drive signal that cancels the known nonlinear response of the electro-optic modulator.
\paragraph{Algorithmic compensation.} Alternatively, the nonlinearity can be accounted for in the signal model~\cite{Isleif2016}. For example, the analytical model of the signal can be expanded to include the nonlinear terms $\epsilon, \psi_2$. An extended readout algorithm then estimates the distortion simultaneously with the primary parameters, effectively separating its influence from the length measurement.

\subsection{Residual amplitude modulation}

\begin{figure*}[htpb]
\centering
\includegraphics{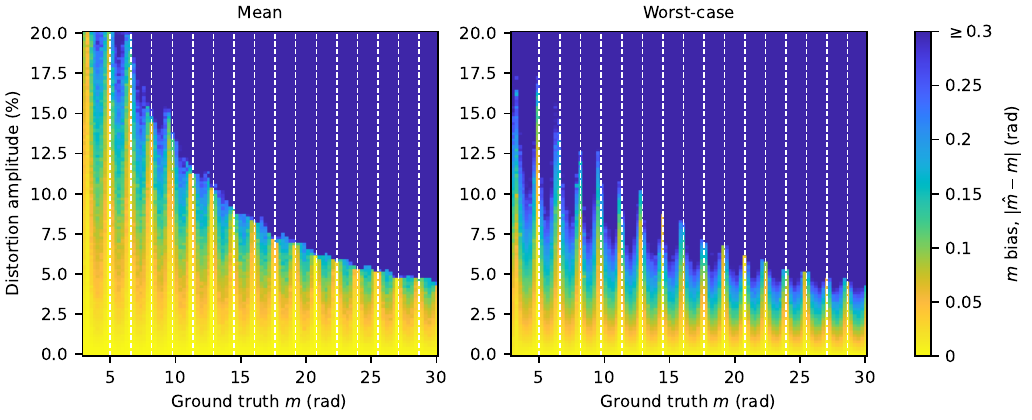}
\includegraphics{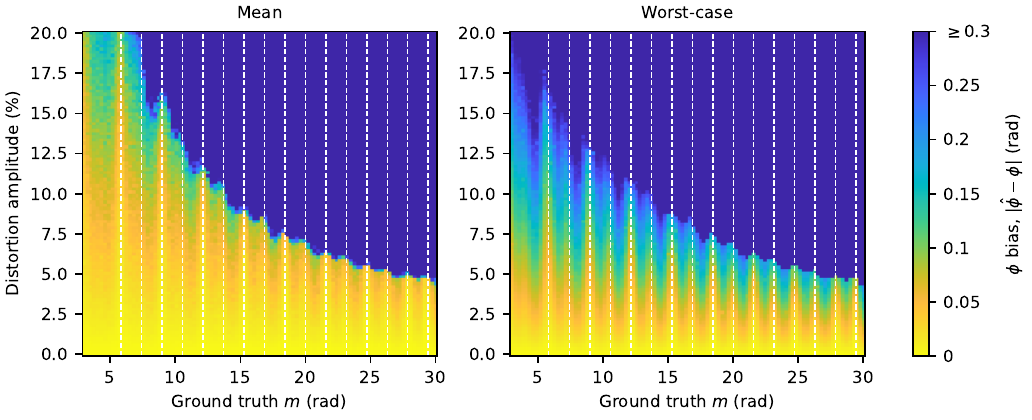}
\caption{Bias in the estimated modulation depth ($\hat{m}$) and interferometric phase ($\hat \phi$) due to residual amplitude modulation. The bias is mapped as a function of the true modulation depth $m$ and the RAM amplitude $\epsilon_{\rm AM}$, with each pixel representing the result of a Monte Carlo simulation over the unknown phases $\Phi$ and $\psi_{\rm AM}$. The~(left) panels show the mean bias, while the (right) panels show the worst-case bias. The plot reveals distinct vertical ``valleys of robustness'' where the bias is strongly suppressed. The $\hat m$ valleys align with the extrema of $J_1(2m)$, i.e., occurring at $m \approx 4.27, 5.85, 7.43, \dots$, and the $\hat \phi$ valleys align with the zeros of $J_1(2m)$, i.e., occurring at $m \approx 3.51, 5.09, 6.66 \dots$. The dashed white lines indicate the theoretical predictions.}
\label{fig:ram_bias}
\end{figure*}

Residual amplitude modulation (RAM) is another common systematic error, occurring when the frequency modulation process spuriously modulates the laser amplitude. We model this by allowing the signal amplitude to be time-dependent:
\begin{equation}
    C(t) = C_0 \left[1 + \epsilon_{\rm AM} \cos(2 \pi f_m t + \psi_{\rm AM})\right],
\end{equation}
where $\epsilon_{\rm AM} \ll 1$ is the fractional RAM amplitude. The total measured signal is therefore the ideal interferometric signal multiplied by this time-varying amplitude. The resulting perturbation signal, $\delta v(t)$, is
\begin{align}
\delta v(t) = C_0 \epsilon_{\rm AM} &\cos(2 \pi f_m t + \psi_{\rm AM}) \nonumber \\
\cdot &\cos(\Phi + m\cos(2 \pi f_m t + \psi)).
\end{align}
This multiplicative error at the fundamental modulation frequency, $f_m$, causes spectral mixing between adjacent harmonics ($n \pm 1$). This is a distinct physical mechanism from the $2 f_m$ phase perturbation of second-harmonic distortion, which mixes harmonics separated by two indices ($n \pm 2$), and leads to a different pattern of bias.

As with modulation nonlinearity, the bias from RAM is not uniform with $m$. Simulations in Figure~\ref{fig:ram_bias} reveal ``valleys of robustness'' where the bias is strongly suppressed. These can be predicted by applying the same time-domain orthogonality principle discussed previously. The condition that the perturbation $\delta v(t)$ be orthogonal to the signal gradients, $\partial v/\partial m$ and $\partial v/\partial \Phi$, leads to specific requirements on the modulation depth (derived in Appendix~\ref{appendix:orthogonality_derivation}).
\begin{itemize}
    \item For the modulation depth estimate $\hat{m}$, the bias is minimized when $J'_1(2m) = 0$. This correctly predicts robustness valleys at the extrema of $J_1(2m)$, such as $m \approx 4.27, 5.85, \dots$.
    \item For the interferometric phase estimate $\hat \phi$, the bias is minimized when $J_1(2m) = 0$. This predicts valleys at the roots of $J_1(2m)$, such as $m \approx 3.51, 5.09, \dots$.
\end{itemize}
These analytical predictions show excellent agreement with the simulated results in Figure~\ref{fig:ram_bias}, confirming that operating at a specific modulation depth is a powerful strategy for passively mitigating RAM.

\subsubsection*{Mitigation strategies}

Beyond selecting robust operating points, RAM can be actively mitigated at both the hardware and software levels.
\paragraph{Hardware mitigation.} The most direct approach is to implement an active power stabilization feedback loop~\cite{Isleif2019}. A photodiode monitors a fraction of the laser power, and a servo controller adjusts the laser drive current to suppress power fluctuations at the modulation frequency.
\paragraph{Algorithmic mitigation.} Alternatively, the effect of RAM can be corrected in software. One common technique is to normalize the AC signal by the measured DC component, which can cancel the time-dependent amplitude~\cite{An2024}. A more general and robust approach is to extend the state-space model of the readout algorithm to include the RAM parameters ($\epsilon_{\rm AM}, \psi_{\rm AM}$). The NLS or EKF algorithm then estimates these parameters simultaneously with the interferometric parameters, effectively separating the RAM-induced error from the length measurement.

\subsection{Ghost beams}
\label{sec:ghost_beam}

\begin{figure*}
\centering
\includegraphics{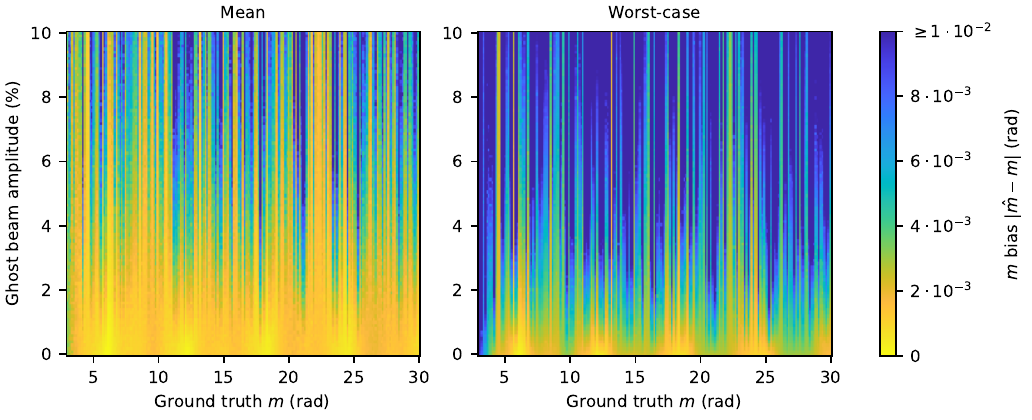}
\includegraphics{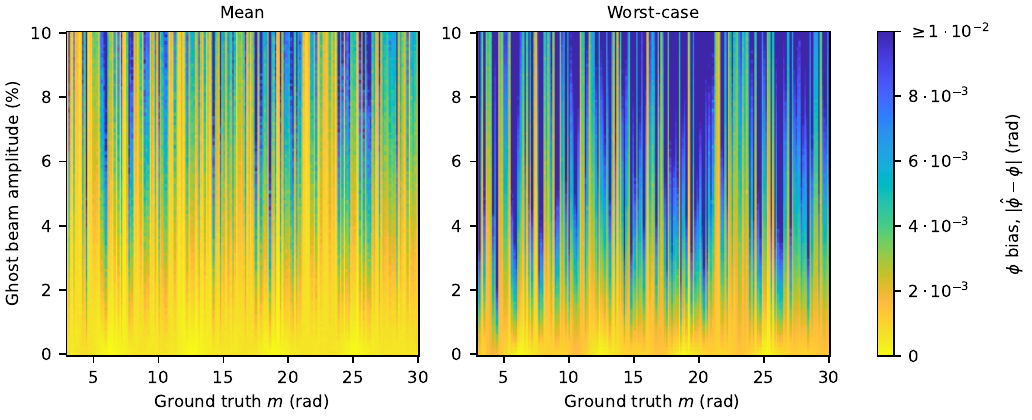}
\caption{Bias in the estimated modulation depth ($\hat{m}$) and interferometric phase ($\hat \phi$) caused by a single parasitic ghost beam. The simulation sweeps the true modulation depth ($m$) and the relative ghost beam amplitude. For each point, a Monte Carlo analysis is performed over the unknown interferometric and ghost beam phases to find the mean and worst-case bias. The plot reveals distinct vertical ``valleys of robustness'' which appear to be distributed seemingly unpredictably.}
\label{fig:ghost_beams}
\end{figure*}

Parasitic reflections from optical surfaces can generate unwanted ``ghost beams'' that interfere with the primary reference and measurement beams. This creates spurious interferometric signals that corrupt the measurement, a common source of periodic error in precision interferometry~\cite{Fleddermann2018}. In DFMI, this problem is particularly nuanced because a ghost beam creates a parasitic DFMI signal with its own distinct modulation depth, $m_{\rm ghost}$, which superimposes onto the desired signal.

We model the total AC voltage signal as the sum of the nominal interferogram and the two primary beat notes generated by a single ghost beam of fractional amplitude $\epsilon_g$ and phase $\Phi_g$~\cite{Gerberding2021}:
\begin{multline}
    v_{\rm AC}(t) = C \Big[ \cos(\Phi + m\cos(2 \pi f_m t)) \\
    + \epsilon_g \cos(\Phi_g + m_g\cos(2 \pi f_m t)) \\
    + \epsilon_g \cos((\Phi - \Phi_g) + (m-m_g)\cos(2 \pi f_m t)) \Big].
\end{multline}
This model shows the superposition of three distinct DFMI signals with modulation depths $m$, $m_g$, and $m-m_g$.

Numerical simulations of this effect (Figure~\ref{fig:ghost_beams}) reveal a complex bias structure. The bias in $\hat{m}$ and $\hat \phi$ scales with the ghost amplitude $\epsilon_g$ and exhibits a strong, non-uniform dependence on the true modulation depth $m$. ``Valleys of robustness'' appear, but unlike the periodic valleys seen for other error sources, their locations are dense and seemingly unpredictable. This complexity arises because the condition for signal orthogonality now involves a complicated sum over products of different Bessel functions ($J_n(m)$, $J_n(m_g)$, $J_n(m-m_g)$), whose zeros do not follow a simple pattern. Nonetheless, the simulations show a notable clustering of these valleys around values of $m \approx 2\pi k$ for integers $k$, which may suggest an underlying spectral symmetry of the primary signal at these operating points that makes the readout algorithm particularly robust against the less-structured parasitic signals.

Despite this complexity, ghost beam errors are largely correctable. Mitigation strategies, detailed by Gerberding and Isleif~\cite{Gerberding2021}, combine optical and algorithmic approaches. Careful optical design (e.g., using wedged optics and anti-reflection coatings) can minimize the generation of ghost beams. Techniques like balanced detection can optically cancel specific parasitic interference terms. Most powerfully, the readout algorithm itself can be extended to suppress their influence. Because the parasitic signal has a different modulation depth, it is mathematically orthogonal to the primary signal. By augmenting the NLS model to simultaneously fit for both the primary signal and a parasitic signal with a different modulation depth, the ghost beam's influence can be effectively isolated and removed from the primary parameter estimates, reducing the error by orders of magnitude~\cite{Gerberding2021}.

\subsection{Scale factor calibration errors}
\label{subsec:cal_error}

Absolute length reconstruction relies on accurately calibrated scale factors. Errors in the estimated frequency modulation amplitude, $\Delta f_{\rm est}$, and the laser carrier frequency, $f_{0, \rm est}$, propagate directly into a bias in the coarse phase estimate, $\hat \Phi_{\text{coarse}}$.

The calibration of $\Delta f_{\rm est}$ is typically a dominant source of systematic uncertainty, with relative uncertainties often limited to the order of 10 parts-per-million (ppm). The laser carrier frequency $f_{0, \rm est}$, in contrast, can be measured with commercial wavemeters to a relative uncertainty of 1 ppm or better, making its contribution to the error budget smaller.

Let the true values be denoted $f_0$ and $\Delta f$, and the calibrated estimates be $\Delta f_{\rm est} = \Delta f + \delta_{\Delta f}$ and $f_{0, \rm est} = f_0 + \delta_{f_0}$, where $\delta_x$ is the absolute calibration error. The coarse phase estimate is $\hat{\Phi}_{\text{coarse}} = \hat{m} (f_{0, \rm est} / \Delta f_{\rm est})$. Its deviation from the true total phase, $\Phi = m(f_0/\Delta f)$, has both a systematic and a statistical component. The systematic bias, $b_{\rm coarse}$, is driven by the calibration errors, while the statistical uncertainty, $\sigma_{\rm coarse}$, is driven by the random error in the measurement of the modulation depth, $\sigma_m$. To first order, these are:
\begin{align}
    b_{\rm coarse} &=  \mathbb{E}\left[ \hat \Phi_{\rm coarse} \right] - \Phi \approx \Phi \left( \frac{\delta_{f_0}}{f_0} - \frac{\delta_{\Delta f}}{\Delta f} \right) \label{eq:Bcal} \\
    \sigma_{\rm coarse} &= \sqrt{\mathbb{E}[(\hat{\Phi}_{\rm coarse} - \mathbb{E}[\hat{\Phi}_{\rm coarse}])^2]} = \left|\frac{f_{0, \rm est}}{\Delta f_{\rm est}}\right| \sigma_m \label{eq:phi_coarse_stat}
\end{align}

For reliable ambiguity resolution, the total error in the coarse phase must be less than half a fringe. A robust design requires this to hold even in the worst-case scenario, leading to the condition $|b_{\rm coarse}| + K \cdot \sigma_{\rm coarse} < \pi$, where $K$ is a confidence factor.

Substituting the expressions for bias and uncertainty, this condition becomes:
\begin{equation}
    \left|\frac{f_0}{\Delta f}\right| \left( \left|m \left( \frac{\delta_{f_0}}{f_0} - \frac{\delta_{\Delta f}}{\Delta f} \right) \right| + K \sigma_m \right) < \pi.
    \label{eq:phi_coarse_error_final}
\end{equation}
Neglecting the smaller contribution from the $f_0$ calibration error simplifies this further. This expression reveals the fundamental trade-offs: the large ratio $|f_0/\Delta f|$ acts as a gain factor, amplifying both the systematic calibration error (which scales with $m$) and the statistical measurement error ($\sigma_m$).

This challenge is particularly acute in long-baseline interferometry where the total phase $\Phi$ is large. The total error in the ambiguity resolution step scales linearly with the optical path difference, $\Delta l$. This can be seen by considering two common design constraints: if $m$ is held constant (requiring $\Delta f \propto 1/\Delta l$), the gain factor $|f_0/\Delta f|$ grows with $\Delta l$. If $\Delta f$ is held constant (e.g., due to hardware limits), then $m \propto \Delta l$, causing the systematic term in Equation~\eqref{eq:phi_coarse_error_final} to grow with $\Delta l$. In either case, the requirements for calibration accuracy and measurement precision become proportionally more stringent as the baseline length increases.

\section{Results: Unified Error Budget and Design Trade-offs}
\label{sec:results}

This section synthesizes the preceding analyses into a unified error budget to guide practical system design. The central challenge is ambiguity resolution, where the total error in the coarse phase estimate must remain below $\pi$. This error budget is overwhelmingly dominated by biases affecting the modulation depth estimate, $\hat{m}$, as any such bias is amplified by the large gain factor $f_0/\Delta f$ (typically $10^4 - 10^5$). Among these biases, modulation nonlinearity constitutes the most significant and difficult-to-correct systematic contribution.

\begin{figure}
\centering
\includegraphics[width=\columnwidth]{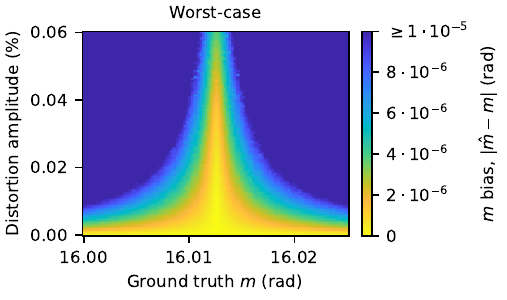} 
\caption{Simulated worst-case bias in the estimated modulation depth ($\hat{m}$) in the vicinity of the $m=16$ robustness valley, for a state-of-the-art arbitrary waveform generator.}
\label{fig:distortion_final}
\end{figure}

To ground our analysis, we establish a ``Baseline'' performance scenario that assumes aggressive mitigation: the use of a high-purity signal source, such as a commercial arbitrary waveform generator with harmonic distortion below -70\,dBc~\cite{AWG}, combined with operation at a ``robustness valley'' (e.g., near $m=16$, see Figure~\ref{fig:distortion_final}). This yields a budgeted residual bias on the modulation depth estimate, which we denote $b_m$, of $10\,\upmu$rad. We also consider ``Degraded'' scenarios with larger residual biases to explore the impact of this critical parameter.

\begin{table}[htpb!]
\caption{Parameters used for generating the three-panel performance heatmap in Figure~\ref{fig:performance_heatmap}. The table lists parameters common to all three scenarios, followed by the specific values for each panel.}
\label{tab:heatmap_params}
\centering
\begin{tabular}{l r}
\toprule
\textbf{Parameter} & \textbf{Value} \\
\midrule
\multicolumn{2}{l}{\textit{\textbf{Common Parameters}}} \\
\quad Laser wavelength, $\lambda_0$ & $1064\,$nm \\
\quad Laser frequency, $f_0$ & $281.76\,$THz \\
\quad Sampling frequency, $f_s$ & $200\,$kHz \\
\quad Effective modulation depth, $m$ & $16\,$rad \\
\quad Laser frequency noise at 1\,Hz, $A_{\rm f, RW}~~~~$ & $100\,\mathrm{kHz}/\sqrt{\mathrm{Hz}}$ \\
\quad Budgeted $\hat{m}$ bias, $b_m$ & $10\,\upmu {\rm rad}$ \\
\quad Statistical confidence factor, $K$ & $3$ \\
\midrule
\multicolumn{2}{l}{\textit{\textbf{Scenario 1: Left Panel}}} \\
\quad Optical path difference, $\Delta l$ & $5\,$cm \\
\quad Modulation amplitude, $\Delta f$ & $15.32\,$GHz \\
\quad Gain factor, $f_0/\Delta f$ & $1.84 \times 10^4$ \\
\midrule
\multicolumn{2}{l}{\textit{\textbf{Scenario 2: Center Panel}}} \\
\quad Optical path difference, $\Delta l$ & $20\,$cm \\
\quad Modulation amplitude, $\Delta f$ & $3.83\,$GHz \\
\quad Gain factor, $f_0/\Delta f$ & $7.36 \times 10^4$ \\
\midrule
\multicolumn{2}{l}{\textit{\textbf{Scenario 3: Right Panel}}} \\
\quad Optical path difference, $\Delta l$ & $50\,$cm \\
\quad Modulation amplitude, $\Delta f$ & $1.53\,$GHz \\
\quad Gain factor, $f_0/\Delta f$ & $1.84 \times 10^5$ \\
\bottomrule
\end{tabular}
\end{table}

\begin{figure*}
\centering
\includegraphics[width=\textwidth]{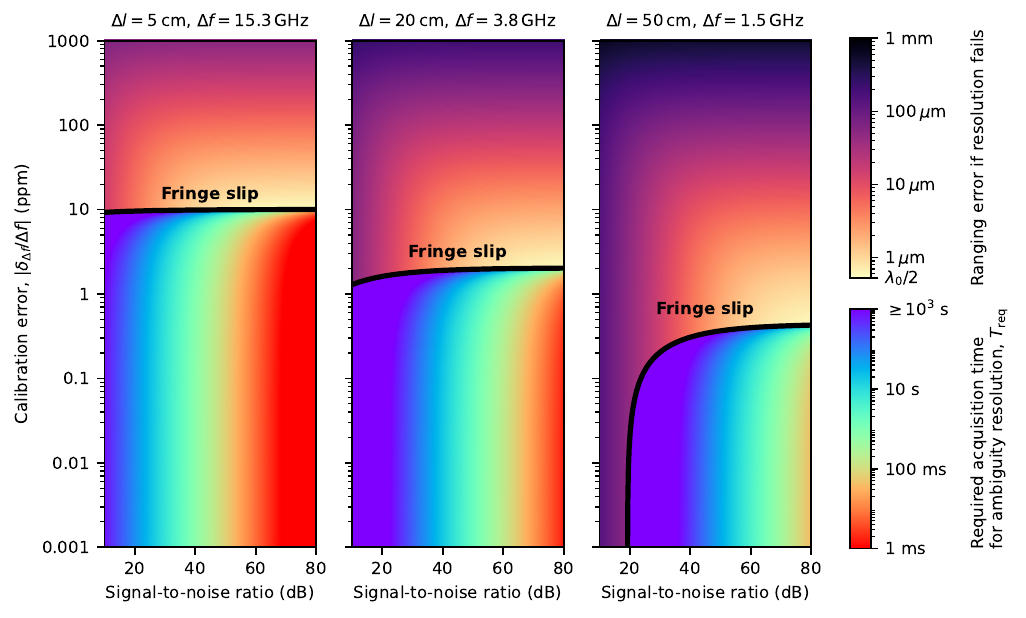} 
\caption{Operational space for absolute length determination for three interferometer baselines ($\Delta l$), each designed to operate in a robustness valley at $m=16$. The lower region shows the required acquisition time for successful ambiguity resolution ($K=3$), while the upper region shows the coarse length error bound when resolution fails. The black curve marks the ``fringe-slip'' boundary where no solution exists. This boundary is the lower envelope of a flat systematic limit and a curved random-noise limit, with the curvature caused by laser frequency drift. Longer baselines increase the impact of calibration error and laser drift, dramatically shrinking the operational space.}
\label{fig:performance_heatmap}
\end{figure*}

\begin{figure*}
\centering
\includegraphics[width=0.8\textwidth]{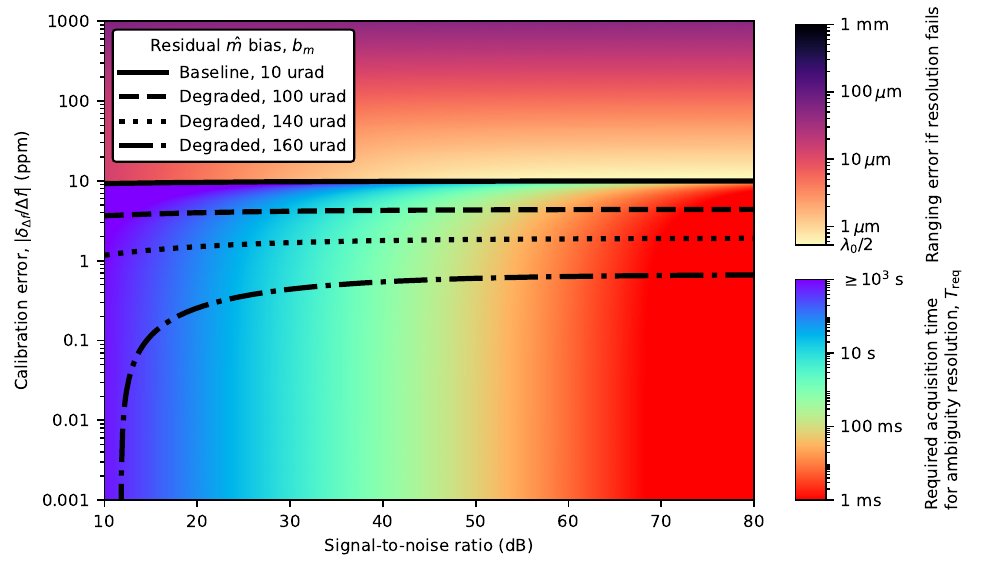}
\caption{Sensitivity of the operational space to systematic error mitigation for the 5\,cm baseline scenario. The fringe-slip boundary is shown for four different assumptions for the residual bias on $\hat{m}$. The ``Baseline'' curve assumes a high-performance system ($10\,\upmu$rad residual bias), while the other curves show how the space progressively shrinks as this bias increases, demanding significantly better calibration to avoid a fringe slip.}
\label{fig:sensitivity_heatmap}
\end{figure*}

The comprehensive ambiguity resolution condition requires that the sum of all biases and random errors on the coarse phase remains below the $\pi$ threshold. The total systematic bias is the sum of the contribution from scale factor calibration, $|b_{\rm cal}|$ [Equation~\eqref{eq:Bcal}], and the amplified residual nonlinearity bias, $|b_{\rm NL}| = |b_m| (f_0/\Delta f)$. The total random error is dominated by the quadrature sum of the amplified statistical uncertainty of the $\hat{m}$ estimate, $\sigma_{\rm coarse}$ [Equations~\eqref{eq:CRLB_voltage} and~\eqref{eq:phi_coarse_stat}], and the integrated laser frequency drift, $\sigma_{\rm drift}$ [Equation~\eqref{eq:phase_noise_random_walk}]. The complete error budget is therefore:
\begin{equation}
    |b_{\rm cal}| + |b_m|\frac{f_0}{\Delta f} + K \sqrt{\sigma_{\rm coarse}^2(T_{\rm acq}) + \sigma_{\rm drift}^2(T_{\rm acq})} < \pi.
    \label{eq:unified_budget_full}
\end{equation}
The two random error terms have opposing dependencies on acquisition time, $T_{\rm acq}$: statistical uncertainty ($\sigma_{\rm coarse}^2 \propto 1/T_{\rm acq}$) decreases with time, while integrated drift noise ($\sigma_{\rm drift}^2 \propto T_{\rm acq}$) increases. This competition implies that a solution for the required acquisition time, if one exists, may only be found within a finite window, $T_{\rm min} < T_{\rm req} < T_{\rm max}$. A ``fringe slip'' occurs when no real solution for $T_{\rm req}$ exists. The boundary of the operational space shown in the following figures is analytically defined by finding the point where the solution to Equation~\eqref{eq:unified_budget_full} for the acquisition time ceases to be real.

Figure~\ref{fig:performance_heatmap} visualizes this complete operational space for the parameters listed in Table~\ref{tab:heatmap_params}. It illustrates the trade-off between the required signal quality (SNR), acquisition time, and calibration fidelity needed to achieve successful ambiguity resolution. A clear trend emerges: as the interferometer baseline $\Delta l$ increases, the operational space shrinks dramatically. For the 50\,cm baseline, success is confined to a narrow region demanding a minimal combination of SNR and acquisition time, as well as exceptionally low calibration uncertainty, highlighting a key challenge for long-baseline DFMI.

This analysis reveals that the most effective way to expand the operational space is to reduce the total systematic bias. Figure~\ref{fig:sensitivity_heatmap} quantifies this by showing the fringe-slip boundary for different levels of the residual bias on $\hat{m}$. The effect is dramatic: increasing $b_m$ from the baseline $10\,\upmu$rad to $160\,\upmu$rad shrinks the operational space by more than an order of magnitude. This provides a direct, quantitative link between the quality of systematic error correction and the achievable performance of the absolute metrology system, underscoring that for high-accuracy applications, engineering effort is best focused on mitigating hardware imperfections (improving $b_m$), and achieving high-quality calibration estimates (improving $b_{\rm cal}$).

\section{Summary and Outlook} \label{sec:conclusion}

In this paper, we have established a comprehensive framework for quantifying the precision and accuracy of absolute length determination via DFMI, by analyzing the dominant random (Section~\ref{sec:Statistical_Precision}) and systematic (Section~\ref{sec:Systematic_Limitations}) error sources.

We began by deriving the Cram\'er-Rao Lower Bounds for $m$ and $\phi$, and showing that both NLS and EKF estimators achieve these bounds (Figure~\ref{fig:crlb-nls-ekf}). The central contribution of this work, however, is the detailed characterization of systematic biases arising from hardware imperfections and scale-factor calibration errors. We developed a digital-twin model of a DFMI experiment using the \texttt{DeepFMKit} toolkit~\cite{Dovale2025}, and systematically analyzed errors arising from modulation nonlinearity (Figure~\ref{fig:distortion_m}), residual amplitude modulation (Figure~\ref{fig:ram_bias}), and ghost beams (Figure~\ref{fig:ghost_beams}) via extensive Monte Carlo simulations.

This analysis revealed the existence of specific operating points where biases are strongly suppressed. For the case of second harmonic distortion and residual amplitude modulation, we introduced a theoretical framework based on signal orthogonality that explains the origin and predicts the locations of these points.

This analysis further revealed that the specific modulation-depth values that suppress bias from different systematic sources are governed by the zeros of distinct Bessel functions or their derivatives. This results in a structured yet interlaced landscape of optimal operating points (Figure~\ref{fig:valleys-of-robustness}). Each robustness valley corresponds to a different error mechanism and parameter being estimated, and no single modulation depth $m$ simultaneously minimizes all bias types. Consequently, the choice of $m$ must be a deliberate trade-off informed by the instrument's specific error budget and the dominant imperfections expected in its implementation.

The bias landscape maps derived in this work also provide a quantitative tool for prioritizing engineering effort. \textcolor{black}{The ordering discussed here refers to unmitigated bias magnitudes in the baseline signal model: under those assumptions, second-harmonic distortion produces the largest raw bias, typically exceeding the residual amplitude modulation and ghost-beam terms by about an order of magnitude. In practical systems, however, the residual limiting factor depends on the mitigation strategy. For example, previous work has shown that RAM and ghost-beam contributions can be substantially reduced through a combination of passive and active suppression methods, including laser power stabilization, optical design, and extended fit models~\cite{Isleif2019, Gerberding2021}.}

We then synthesized these findings into a consolidated error budget for absolute length determination (Figures~\ref{fig:performance_heatmap} and~\ref{fig:sensitivity_heatmap}). This framework clarifies the trade-offs between statistical noise, laser frequency drift, and systematic biases from instrument calibration and hardware imperfections. A key insight is that for long-baseline systems, the primary engineering challenge shifts from overcoming statistical limits (e.g., via longer integration or improved signal-to-noise ratio) to conquering systematic limits through high-fidelity calibration and mitigation of hardware imperfections.

This work points to two clear directions for future advancement: First, the observed robustness valleys invite experimental verification. Controlled DFMI measurements with tunable modulation depths and calibrated sources of distortion can confirm the numerically demonstrated suppression of systematic biases. Such experiments would not only validate the analytical and numerical framework presented here but also establish quantitative benchmarks for optimizing DFMI instruments in practice. Second, the need for ``self-calibrating'' readout algorithms. By augmenting the NLS or EKF state-space to include parameters for these distortion effects, one could identify and dynamically cancel these effects.

The combination of passive robustness and algorithmically enforced active correction represents a new frontier for absolute length metrology with DFMI. Together, these approaches promise DFMI architectures capable of achieving sub-wavelength accuracy, approaching the fundamental precision limits defined in this work.

\section{References}

%


\section*{Acknowledgements}

The author gratefully acknowledges Gerhard Heinzel and Felipe Guzm\'an for their foundational role in developing Deep Phase Modulation Interferometry, which served as the technical and conceptual precursor to Deep Frequency Modulation Interferometry (DFMI). Oliver Gerberding and Katharina-Sophie Isleif are acknowledged for their pioneering contributions in DFMI. The author also thanks Victor Huarcaya for his extensive work on the experimental implementation of DFMI, and Stefano Gozzo and Reshma Krishnan Sudha for many insightful discussions regarding the NLS fitting algorithm.


\appendix

\section{Error from the first-order Taylor approximation}
\label{appendix:taylor_error}

The DFMI signal model in Section~\ref{sec:DFMI_Signal} uses a first-order Taylor approximation for the differential phase modulation. The exact expression is
\begin{equation}
\Delta\phi_{\text{mod}}(t) = \phi_{\text{mod}}(t) - \phi_{\text{mod}}(t-\tau).
\label{eq:true-mod}
\end{equation}
This appendix provides a formal analysis of the higher-order terms neglected in the approximation.

Expanding $\phi_{\text{mod}}(t-\tau)$ as a Taylor series around $t$ yields:
\begin{equation}
\Delta\phi_{\text{mod}}(t) = \tau\phi'_{\text{mod}}(t) - \frac{\tau^2}{2}\phi''_{\text{mod}}(t) + \frac{\tau^3}{6}\phi'''_{\text{mod}}(t) - \mathcal{O}(\tau^4).
\end{equation}
Substituting the derivatives of $\phi_{\text{mod}}(t) = (\Delta f/f_m) \sin(2 \pi f_m t + \psi)$ and using the definition of the modulation depth, $m = 2\pi\Delta f \tau$, allows us to express the full differential phase modulation as:
\begin{widetext}
\begin{equation}
\Delta\phi_{\text{mod}}(t) \approx m \cos(2 \pi f_m t + \psi) + \underbrace{\frac{m (2 \pi f_m \tau)}{2} \sin(2 \pi f_m t + \psi)}_{\text{Quadrature Distortion}} - \underbrace{\frac{m (2 \pi f_m \tau)^2}{6} \cos(2 \pi f_m t + \psi)}_{\text{Amplitude Bias}}.
\end{equation}
\end{widetext}
This reveals two error mechanisms introduced by the approximation: an amplitude bias, which acts as a small fractional change to $m$ of magnitude $(2 \pi f_m \tau)^2 / 6$, and a quadrature distortion term. The quadrature distortion, being first-order in the small parameter $2 \pi f_m \tau$, is the dominant error source.

\begin{figure*}
\centering
\includegraphics[width=\columnwidth]{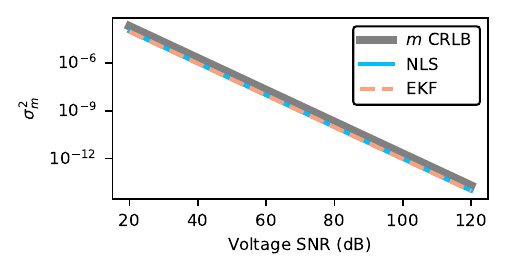}
\includegraphics[width=\columnwidth]{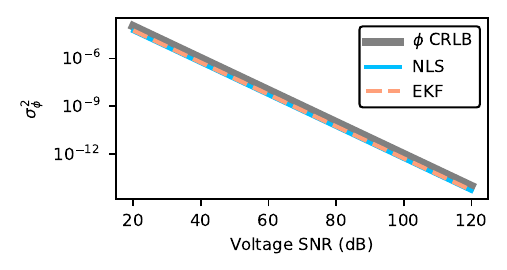}
\caption{Variance of the estimated modulation depth $\hat m$~(left panel) and interferometric phase $\hat \phi$ (right panel) from NLS and EKF readouts on a synthetic signal, as a function of the voltage SNR. Monte Carlo simulations (500 trials per point with additive Gaussian noise and randomized interferometric phase) demonstrate that both estimators approach the analytically derived CRLBs. The estimators process data over one modulation period ($1/f_m = 1\,\rm ms$) sampled at 200\,kHz, corresponding to $N_v=200$ samples.}\label{fig:crlb-nls-ekf}
\end{figure*}

To quantify its impact, consider a demanding long-baseline scenario with $\Delta l = 10\,\rm m$ and $f_m = 1\,\rm kHz$. The expansion parameter is $2 \pi f_m \tau \approx 2.1 \times 10^{-4}$. The fractional amplitude of the quadrature distortion relative to the main modulation depth is $(2 \pi f_m \tau)/2 \approx 105$\,ppm. While not insignificant, this error is typically overshadowed by other systematic effects. For long-baseline systems, the error from scale factor calibration (Section~\ref{subsec:cal_error}) grows linearly with the baseline length and becomes the dominant limitation long before the Taylor approximation error is a major contributor to the budget.

All simulations presented in this work use the exact differential phase from Equation~\eqref{eq:true-mod}, and therefore inherently include the effects of this small perturbation.

\section{CRLB derivations for $m$ and $\Phi$}
\label{sec:CRLB_derivation}

This appendix details the derivation of the asymptotic Cram\'er-Rao Lower Bound (CRLB) for the modulation depth, $m$, and the interferometric phase, $\Phi$.

The CRLB is found from the inverse of the Fisher Information Matrix (FIM). While the full bound requires inverting the complete 4x4 FIM, the asymptotic limit can be found by analyzing the diagonal elements, as the off-diagonal correlation terms become negligible in the asymptotic limit when averaged over all phases. The diagonal FIM element corresponding to the modulation depth $m$ is:
\begin{equation}
[\mathcal{I}(\mathbf{x})]_{m,m} = \frac{1}{\sigma_{h}^2} \sum_{n=1}^{N_h} \left[ \left(\frac{\partial I_n}{\partial m}\right)^2 + \left(\frac{\partial Q_n}{\partial m}\right)^2 \right],
\end{equation}
where $\sigma_{h}^2 = 2\sigma_v^2/N_v$ is the noise variance on each demodulated quadrature component over an acquisition of $N_v$ samples. The sum of the squared partial derivatives of the I/Q components [Equations~\eqref{eq:Iana} and~\eqref{eq:Qana}] with respect to $m$ is
\begin{equation}
\left(\frac{\partial I_n}{\partial m}\right)^2 + \left(\frac{\partial Q_n}{\partial m}\right)^2 = (2C)^2 [J'_n(m)]^2 \cos^2\left(\Phi + n \frac{\pi}{2} \right),
\end{equation}
where $J'_n(m)$ is the derivative of the Bessel function. To obtain a general performance limit, we average over the unknown interferometric phase $\Phi$, using $\langle \cos^2(\cdot) \rangle = 1/2$. The expected FIM element becomes:
\begin{equation}
\mathrm{E}_{\Phi}\left[[\mathcal{I}(\mathbf{x})]_{m,m}\right] = \frac{2C^2}{\sigma_{h}^2} \sum_{n=1}^{N_h} [J'_n(m)]^2.
\end{equation}
In the asymptotic limit where all significant harmonics are included, the sum can be approximated by the identity $\sum_{n=1}^{\infty} [J'_n(m)]^2 \approx 1/4$ for $m \gg 1$~\cite{NIST:DLMF}. The asymptotic information is therefore:
\begin{equation}
\lim_{m, N_h \to \infty} \mathrm{E}_{\Phi}\left[[\mathcal{I}]_{m,m}\right] \approx \frac{2C^2}{\sigma_{h}^2} \left( \frac{1}{4} \right) = \frac{C^2}{2\sigma_{h}^2}.
\end{equation}
The asymptotic variance is the inverse of this information. Substituting $\sigma_{h}^2 = 2\sigma_v^2/N_v$ and using the definition of the raw voltage SNR, $\text{SNR}_{v} = C/\sigma_v$, we arrive at the final expression for the CRLB for $m$:
\begin{equation}
\sigma_m^2 \approx \frac{4}{N_v \cdot \text{SNR}_{v}^2},
\end{equation}
which is Equation~\eqref{eq:CRLB_voltage} in the main text.

For completeness, we also note the exact time-domain CRLB form with explicit phase dependence, as discussed in the supplemental derivation of~\cite{Eckhardt2022}. For a DFM signal written as $P(t)=A[1+\cos(m\sin(\omega_m t+\psi)+\Phi)]$ with additive white Gaussian noise of variance $\sigma^2$, the corresponding bound is
\begin{equation}
\mathrm{Var}(\hat m) \ge \frac{4\sigma^2}{f_s T A^2\left[1-\left(J_0(2m)-J_2(2m)\right)\cos(2\Phi)\right]}.
\label{eq:CRLB_m_exact_phase}
\end{equation}
This expression is exact for that signal model and clarifies the origin of the residual $(m,\Phi)$ dependence in finite-harmonic calculations. In the regime used for Figure~\ref{fig:crlb-nls-ekf}, where $\Phi$ is randomized between trials, phase averaging gives $\langle\cos(2\Phi)\rangle=0$, so Equation~\eqref{eq:CRLB_m_exact_phase} reduces to Equation~\eqref{eq:CRLB_voltage}.

An analogous derivation for the interferometric phase $\Phi$ yields its CRLB:
\begin{equation}
\sigma_{\Phi}^2 \approx \frac{2}{N_v \cdot \text{SNR}_v^2 (1-J_0(m)^2)}.
\label{eq:CRLB_phi_voltage}
\end{equation}
The term $1-J_0(m)^2$ is proportional to the total power in the higher harmonics, confirming the physical intuition that phase information is carried by those components, not the carrier. The precision of $\Phi$ is thus maximized when the modulation depth $m$ is chosen to be a root of the Bessel function $J_0(m)$, which transfers all power out of the carrier and into the information-bearing sidebands.

Numerical simulations in Figure~\ref{fig:crlb-nls-ekf} confirm that both NLS and EKF estimators are asymptotically efficient, converging to these bounds.

\section{Derivation of Orthogonality Conditions for Systematic Biases}
\label{appendix:orthogonality_derivation}

\begin{figure*}
\centering
\includegraphics{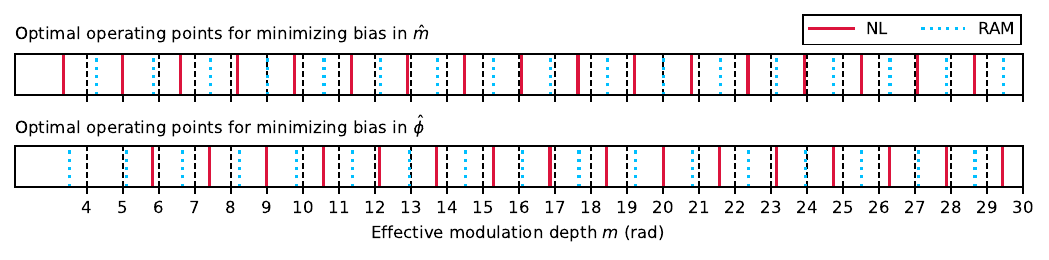}
\caption{Summary of optimal operating points (``valleys of robustness'') for minimizing systematic biases in the estimated modulation depth, $\hat{m}$ (top panel), and interferometric phase, $\hat \phi$ (bottom panel). Vertical lines indicate the $m$-values that provide high intrinsic immunity to biases from either modulation nonlinearity (NL, solid red) or residual amplitude modulation (RAM, dotted blue).}
\label{fig:valleys-of-robustness}
\end{figure*}

This appendix provides the mathematical derivations for the ``valleys of robustness'' conditions presented in Section~\ref{sec:Systematic_Limitations}. The principle is that a parameter estimate is robust against a perturbation, $\delta v(t)$, when that perturbation is orthogonal to the ideal signal's sensitivity to that parameter, $g_x(t) = \partial v_{\rm ideal}/\partial x$. The condition for minimal bias, averaged over all unknown phases, is $\langle \langle \delta v(t), g_x(t) \rangle \rangle_{\text{phases}} = 0$.

The evaluation of the inner product integral relies on the identity:
\begin{equation}
\int_{0}^{2\pi} \cos(n\theta - \phi) e^{iZ\cos\theta} \,d\theta = 2\pi i^n J_n(Z) \cos\phi.
\label{eq:bessel_integral_identity}
\end{equation}

\subsection{Modulation Nonlinearity}
For second-harmonic distortion, the perturbation and gradients are:
\begin{align*}
\delta v(t) &\approx -C m \epsilon \cos(2\theta - (2\psi-\psi_2)) \sin(\Phi + m\cos\theta) \\
g_m(t) &= -C \cos(\theta) \sin(\Phi + m\cos\theta) \\
g_\Phi(t) &= -C \sin(\Phi + m\cos\theta),
\end{align*}
where $\theta = 2 \pi f_m t + \psi$.

\subsubsection{Bias in effective modulation depth}
The inner product $\langle \delta v, g_m \rangle$ is proportional to an integral involving $\cos(2\theta - \phi_A)\cos(\theta)\sin^2(\Phi+m\cos\theta)$, where $\phi_A = 2\psi - \psi_2$. Using trigonometric identities and averaging over the phases, the non-zero term that must vanish is the integral
\begin{equation*}
\int_0^{2\pi} \left[\cos(\theta - \phi_A) + \cos(3\theta - \phi_A)\right] e^{i2m\cos\theta} \,d\theta.
\end{equation*}
Applying identity \eqref{eq:bessel_integral_identity} to the $n=1$ and $n=3$ terms yields a result proportional to $J_1(2m) - J_3(2m)$. Using the Bessel recurrence relation $J'_n(x) = \frac{1}{2}[J_{n-1}(x) - J_{n+1}(x)]$, the condition for zero bias is:
\begin{equation}
    J_1(2m) - J_3(2m) = 2J'_2(2m) = 0.
\end{equation}

\subsubsection{Bias in interferometric phase}
The inner product $\langle \delta v, g_\Phi \rangle$ requires the evaluation of an integral proportional to $\int \cos(2\theta - \phi_A) e^{i2m\cos\theta} d\theta$. Applying identity \eqref{eq:bessel_integral_identity} with $n=2$ shows the result is proportional to $J_2(2m)$. The condition for zero bias is thus:
\begin{equation}
    J_2(2m) = 0.
\end{equation}

\subsection{Residual Amplitude Modulation}
For RAM, the perturbation and gradients are:
\begin{align*}
\delta v(t) &= C_0 \epsilon_{\rm AM} \cos(\theta - (\psi-\psi_{\rm AM})) \cos(\Phi + m\cos\theta) \\
g_m(t) &= -C_0 \cos(\theta) \sin(\Phi + m\cos\theta) \\
g_\Phi(t) &= -C_0 \sin(\Phi + m\cos\theta).
\end{align*}

\subsubsection{Bias in effective modulation depth}
The inner product $\langle \delta v, g_m \rangle$ involves the term $\sin(2\Phi+2m\cos\theta)$. The relevant integral is proportional to $\int [\cos(2\theta - \phi) + \cos(\phi)] e^{i2m\cos\theta} d\theta$, where $\phi = \psi - \psi_{\rm AM}$. Applying identity \eqref{eq:bessel_integral_identity} for $n=2$ and $n=0$ gives a result proportional to $-J_2(2m) + J_0(2m)$. Using the recurrence relation, the zero-bias condition is:
\begin{equation}
    J_0(2m) - J_2(2m) = 2J'_1(2m) = 0.
\end{equation}

\subsubsection{Bias in interferometric phase}
The inner product $\langle \delta v, g_\Phi \rangle$ also involves the term $\sin(2\Phi+2m\cos\theta)$. The relevant integral is $\int \cos(\theta - \phi) e^{i2m\cos\theta} d\theta$. Applying identity \eqref{eq:bessel_integral_identity} with $n=1$ yields a result proportional to $J_1(2m)$. The condition for zero bias is:
\begin{equation}
    J_1(2m) = 0.
\end{equation}

This analysis provides the theoretical foundation for the ``valleys of robustness'' observed in the simulations. It shows that the specific modulation depth values that suppress bias from different systematic sources are governed by the zeros of different Bessel functions or their derivatives. As summarized in Figure~\ref{fig:valleys-of-robustness}, this creates a structured, predictable, but interlaced pattern of optimal operating points. No single $m$-value is robust to all error types simultaneously. Therefore, the choice of modulation depth must be a deliberate trade-off, informed by an instrument's specific error budget. This framework transforms the mitigation of dominant systematics from a purely hardware-based challenge into a tractable problem of intelligent parameter selection.

\end{document}